# Crystal structure evolution in the van der Waals transition metal trihalides

**Marie Kratochvílová[1], Petr Doležal[1], Dávid Hovančík[1], Jiří Pospíšil[1], Anežka Bendová[1], Michal Dušek[2], Václav Holý[1], Vladimír Sechovský[1]**

[1] Charles University, Faculty of Mathematics and Physics, Department of Condensed Matter Physics, Ke Karlovu 5, 121 16 Prague 2, Czech Republic
[2] Institute of Physics, Czech Academy of Sciences, Na Slovance 2, 182 21 Prague 8, Czech Republic

**Abstract**

Most transition-metal trihalides are dimorphic. The representative chromium-based triad, $CrCl_3$, $CrBr_3$, $CrI_3$, is characterized by the low-temperature phase adopting the trigonal $BiI_3$-type while the structure of the high-temperature phase is monoclinic of $AlCl_3$ type (C2/m). The structural transition between the two crystallographic phases is of the first-order type with large thermal hysteresis in $CrCl_3$ and $CrI_3$. We studied crystal structures and structural phase transitions of vanadium-based counterparts $VCl_3$, $VBr_3$, and $VI_3$ by measuring specific heat, magnetization, and X-ray diffraction as functions of temperature and observed an inverse situation. In these cases, the high-temperature phase has a higher symmetry while the low-temperature structure reveals a lower symmetry. The structural phase transition between them shows no measurable hysteresis in contrast to $CrX_3$. Possible relations of the evolution of the ratio *c/a* of the unit cell parameters, types of crystal structures, and nature of the structural transitions in V and Cr trihalides are discussed.

## 1. Introduction

Van der Waals (vdW) magnetic materials, including the transition-metal (*TM*) trihalides $TMX_3$ ($X$ = Cl, Br, I), have gained immense popularity in recent years after being studied for decades primarily as magnetic model structures in layered compounds. The crystal structure of *TM* trihalides given by a graphene-like honeycomb network of *TM* ions and their magnetic ordering at finite temperatures provide promising opportunities for spintronic device applications [1-5].

$CrI_3$ has been the most extensively studied due to its ferromagnetism [6] with the highest Curie temperature ($T_C$ = 61 K [7]) within the trihalide family. Also $CrBr_3$ exhibits ferromagnetism below $T_C$ (= 34 K) [8] whereas $CrCl_3$ becomes antiferromagnetic (AFM) at low temperatures ($T_N$ = 16.8 K [9]). $CrI_3$ and $CrBr_3$ are the only trihalide ferromagnets listed in Table 2 of the review paper of McGuire [10].

Recent research activities were successful in exploring one more $TMX_3$ ferromagnet, namely $VI_3$ [11-14], which has originally been reported much earlier [6]. Curie temperature of $VI_3$ ($T_C$ = 50 K [11,12]) is somewhat lower than $T_C$ of $CrI_3$. However, the physics of $VI_3$ seems to differ from $CrI_3$ in numerous important aspects.

The $CrX_3$ compounds and probably most of the other *TM* trihalides are dimorphic [10]. The $CrX_3$ $\alpha$-phases adopt the trigonal structure of the $BiI_3$ type (space group R-3) and the $\beta$-phase is monoclinic of the $AlCl_3$ type (C2/m), i.e. the lower symmetry crystallographic phase that is preferred at higher temperatures. When cooled from high temperatures the chromium trihalides undergo a first-order structural phase transition from the $\beta$ to the $\alpha$ phase. The transition takes place near 210, 420, and 240 K in $CrI_3$, $CrBr_3$, and $CrCl_3$, respectively, and exhibits large thermal hysteresis (except $CrBr_3$, where the information is not known, yet) determining the temperature range over which both phases coexist [7, 15]. The transition is probably controlled by interlayer interactions, as the monoclinic structure of the β-phase develops from the trigonal α-phase by successive equal small shifts of the basal plane layers. Twinning and stacking faults were reported developing during the transition upon cooling as the layers rearrange themselves into the $BiI_3$ stacking [7].

Contrary to this behavior the high-temperature phase of $VI_3$ adopts the $BiI_3$-type trigonal structure [11,12, 16] and when cooling below room temperature a structural phase transition to the lower-symmetry monoclinic structure has been observed [12,16] at 79 K. Results of recent synchrotron and neutron experiments offered structural models for $VI_3$ [17] and exhibited a relative easiness of preparation of the recently reported $LiVI_3$ phase. Such material belongs to the insertion compounds which show promising properties in the field of applications including Li-ion batteries and beyond [18].

Similarly, the room-temperature $BiI_3$-type trigonal structure is reported for $VBr_3$ and $VCl_3$ [19-21] which both become AFM at low temperatures [6, 22]. In contrast to $VI_3$, the vanadium atoms are found in two sites in the structure of $VBr_3$; the majority of the vanadium atoms form a honeycomb lattice and a nonnegligible fraction occupies the central (normally) vacant site. This partial occupancy is attributed to the presence of stacking faults, which is significantly higher in $VBr_3$ compared to $VI_3$ [21]. A structural transition of $VBr_3$ at 90.4 K has been recently revealed by measurements of specific heat [21].

The apparent controversy between the structural evolution in $CrX_3$ and $VX_3$ motivated us to extend our X-ray diffraction, specific-heat, and magnetization investigation on $VBr_3$ and $VCl_3$ to investigate the high-temperature (HT) and low-temperature (LT) crystal structures and the character of structural phase transition between them. After these experiments, we can confirm that the HT crystal symmetry of $VCl_3$, $VBr_3$, and $VI_3$ is reduced below certain own characteristic temperature $T_s$. The phase transition between the two structure phases is most probably of the first-order type. Contrary to ferromagnetism in $VI_3$, $VCl_3$ and $VBr_3$ order AFM as demonstrated by specific heat in magnetic fields. No symmetry lowering is observed in $VBr_3$ in connection with the transition to the AFM state. The change in the thermal evolution of the crystal symmetry of the crystal structure in the $TMX_3$ halides between the V and Cr-based compounds will be discussed concerning the variations of structure parameters $c$, $a$, and the $c/a$ ratio.

## 2. Methods

Single crystals of $VBr_3$ were synthesized by chemical vapor transport technique by recrystallization of the powder precursor $VBr_3$ (Thermo Fisher Scientific) sealed in a silica tube under vacuum. The tube was placed in a gradient horizontal tube furnace with the hot end at 480 °C and the cold end at 350 °C for approximately two weeks. We obtained shiny plate-like single crystals with dimensions of several square millimeters fast degrading in air similarly to $VI_3$ [23]. Their composition and purity were checked by microprobe analysis which confirmed the 1:3 molar ratio of V and Br, respectively.

The $VCl_3$ powder (97% purity) was obtained commercially from Sigma-Aldrich. The powder was pressed into pellets for specific heat and magnetization measurements. For the low-temperature X-ray powder diffraction, we attached the powder to the sample holder with a thin layer of Apiezon grease and aligned the flat surface of the sample without pressing it. Initial attempts to grow single crystals of $VCl_3$ chemical vapor transport were not successful so far, showing that their preparation might be not that straightforward compared to the other $VX_3$ compounds.

Concerning the moisture sensibility, we minimized the sample's exposure to the necessary mounting time only in the case of all performed experiments. The specific heat and magnetization were measured using a Quantum Design physical property measurement system at temperatures down to 2 K. The X-ray diffraction measurement from 300 K down to 5 K was performed using the diffractometer D500-HR-4K equipped with the closed cycle He cryostat (101J Cryocooler from ColdEdge), helium chamber, and the Attocube ANR 31 sample holder.

## 3. Results and discussion

Figure 1 summarizes the specific heat and magnetization of the $VBr_3$ single crystal and the $VCl_3$ polycrystal. The specific heat of $VBr_3$ confirms the previously reported emergence of a structural transition $T_s$ = 90 K and AFM transition at $T_N$ = 26.5 K as shown in Figure 1(a). $VCl_3$ polycrystal reveals also a structural transition at $T_s$ = 100 K reported for the first time to our knowledge, and an AFM transition at $T_N$ = 21 K (see Figure 1(b)). The powder character of the pellet sample is responsible for the somewhat smeared transitions. One can see in the inset of Figure 1(a) that while the magnetic transition is shifted by ~2 K to lower temperatures in the 9T magnetic field applied perpendicular to the vdW layers in $VBr_3$, the structural transition seems to be untouched by the magnetic field. The AFM transition shifts slightly to lower temperatures while the structural transition remains conserved in the magnetic field of 9 T in the case of $VCl_3$, too. This observation contrasts sharply with the behavior of the structural transition in the $VI_3$ compound which reveals a shift by ~ 2 K to lower temperatures in a comparable magnetic field applied perpendicularly to the layers [16]. The insets in Figure 1(a) and (b) offer a comparison of the AFM transition in the zero and 9T-magnetic fields.

Figure 1(c) presents the temperature dependence of $\partial M/\partial T$ in $VBr_3$ focusing on the vicinity of the structural transition. The data were measured upon cooling down and warming up to be able to recognize any possible thermal hysteresis. The difference between both curves does not provide any clear answer regarding the hysteresis and thus, the nature of the structural transition in $VBr_3$.

Due to the scattered polycrystalline $VCl_3$ data, no peak or bump revealing the structural transition was observed, unlike the Néel transition, as shown in Figure 1(d). However, the change of slope in the $H/M$(T) data is a clear sign of the $T_s$ presence (see Figure S2 in Supplementary Mat.). A linear Curie-Weiss fit to the $VCl_3$ data was performed above $T_s$ as the structural transition alters the slope of the susceptibility. We have extracted an effective moment of $\mu_{eff}$ = 2.96(15) $\mu_B$/V corresponding to the value of 2.85 $\mu_B$ reported by Starr et al. [22] within error and it is also close to the expected S=1 value. The extrapolated Curie temperature $\theta_C$ from the fit is -63(2) K, which differs from the reported one -30.1 K [22] but still clearly confirms the emergence of AFM order in $VCl_3$.

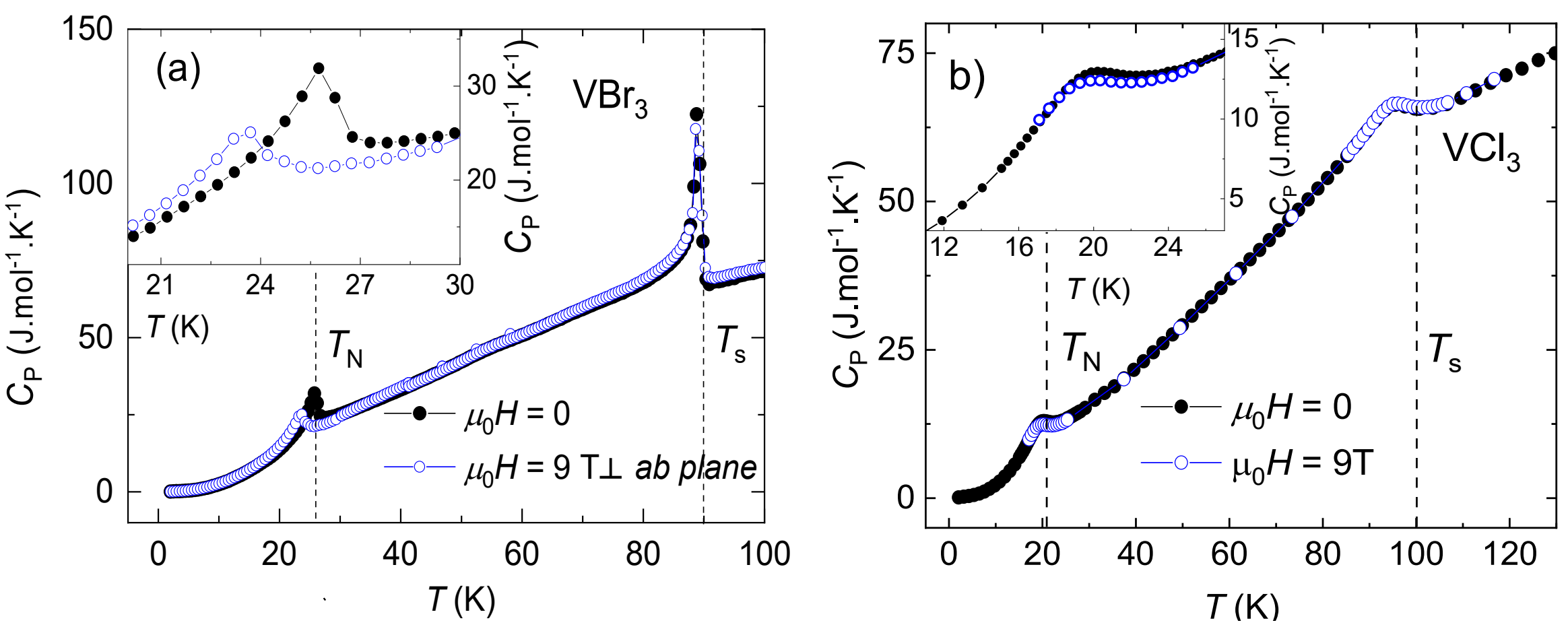

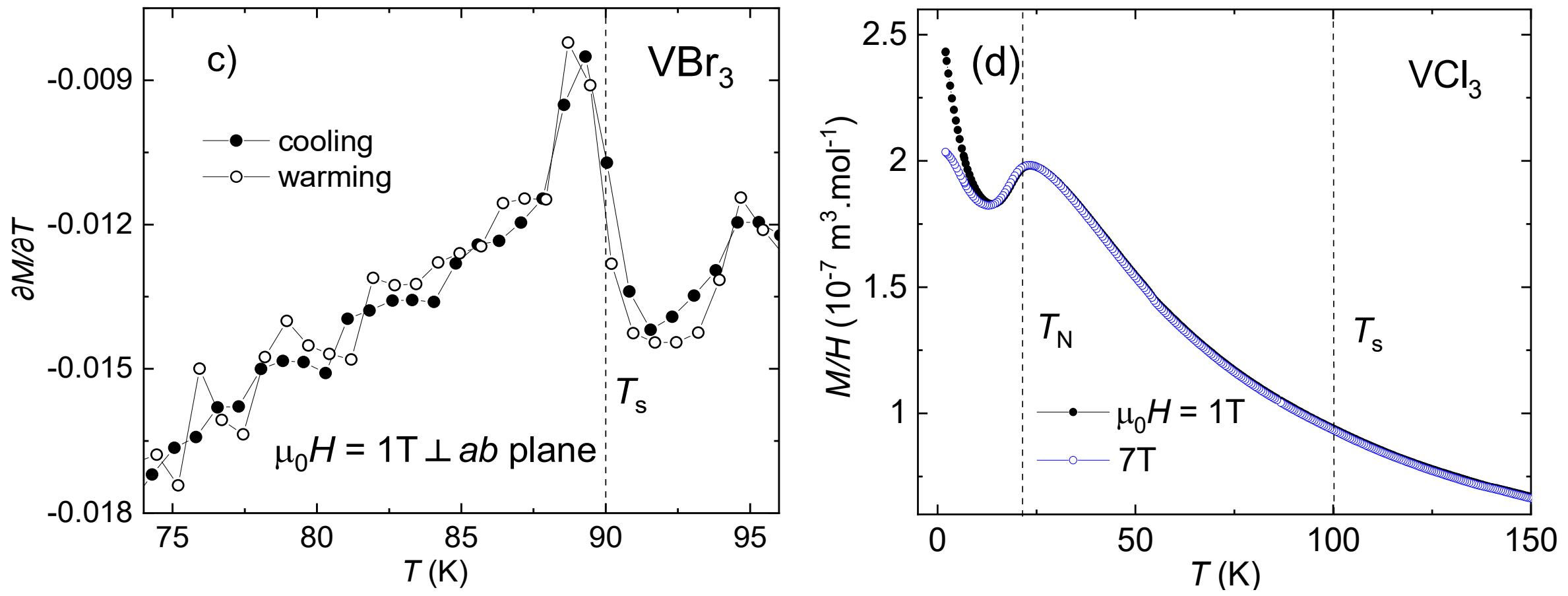

Figure 1. Temperature dependence of the specific heat $C_p$ of (a) $VBr_3$ and (b) $VCl_3$ measured at zero and 9T magnetic field applied perpendicularly to the single-crystal layers (in case of the $VBr_3$ single crystal). The inset shows the detail of the specific heat in the vicinity of the magnetic transition. (c) Temperature dependence of $\partial M/\partial T$ of $VBr_3$ measured in 1T magnetic field applied perpendicularly to the single-crystal layers is depicted in the vicinity of the structural transition measured upon cooling and warming. No measurable hysteresis was detected. (d) Temperature dependence of the magnetization $M/H$ of $VCl_3$ measured in 1T and 7T magnetic fields. The vertical dashed lines mark the positions of the structural and AFM phase transitions at $T_s$ and $T_N$ obtained from the specific heat, respectively.

Turning our attention to the structural analysis, we emphasize that, analogically to the $VI_3$ single crystals, the $VBr_3$ sample shape is a thin (0 0 L) platelet. Together with the geometrical limitations of the low-temperature diffraction setup, only measuring diffraction maxima with L > 0 is possible. Therefore, it is not possible to determine the exact structure of $VBr_3$ by mapping selected diffraction peaks, but the temperature dependence of the lattice parameters can be studied and some predictions about the LT structure can be made based on the splitting of the diffraction peaks.

The hexagonal unit cell for the trigonal crystal system and the corresponding H K L indices were considered according to previous studies [19-21]. The reciprocal space maps were measured around the symmetric (0 0 12), (0 0 15), (0 0 18), (0 0 21), and asymmetric (1 1 21), (-1 2 21), (-2 4 15), (3 3 12) diffraction peaks. The temperature dependence of the asymmetric diffraction peak is shown in Figure 2, which confirms that the crystal symmetry lowers from a trigonal symmetry to a presumably monoclinic one by the structural transition at $T_s$ = 90 K. Below 90 K, the (1 1 L) and (-1 2 L) diffraction peaks are both split into two peaks; both split pairs have different $2\theta$ distances. The diffraction-peak splitting with two diffraction peaks moving to larger and the other to smaller $2\theta$ values, respectively, is ascribed to the reduction of the lattice symmetry and it can be understood by opposite deformation of non-equivalent domains in analogy to the $VI_3$ compound [16]. Noticeably, we can see a coexistence of the diffraction peak belonging to the HT structure phase and of the pairs of diffraction peaks from the LT one at 89.5 K (upper panel in Figure 2). Such effect can be either caused by the first-order character of the structural transition or by phase non-homogeneity of the sample. The nature of the transition will be discussed later in the text.

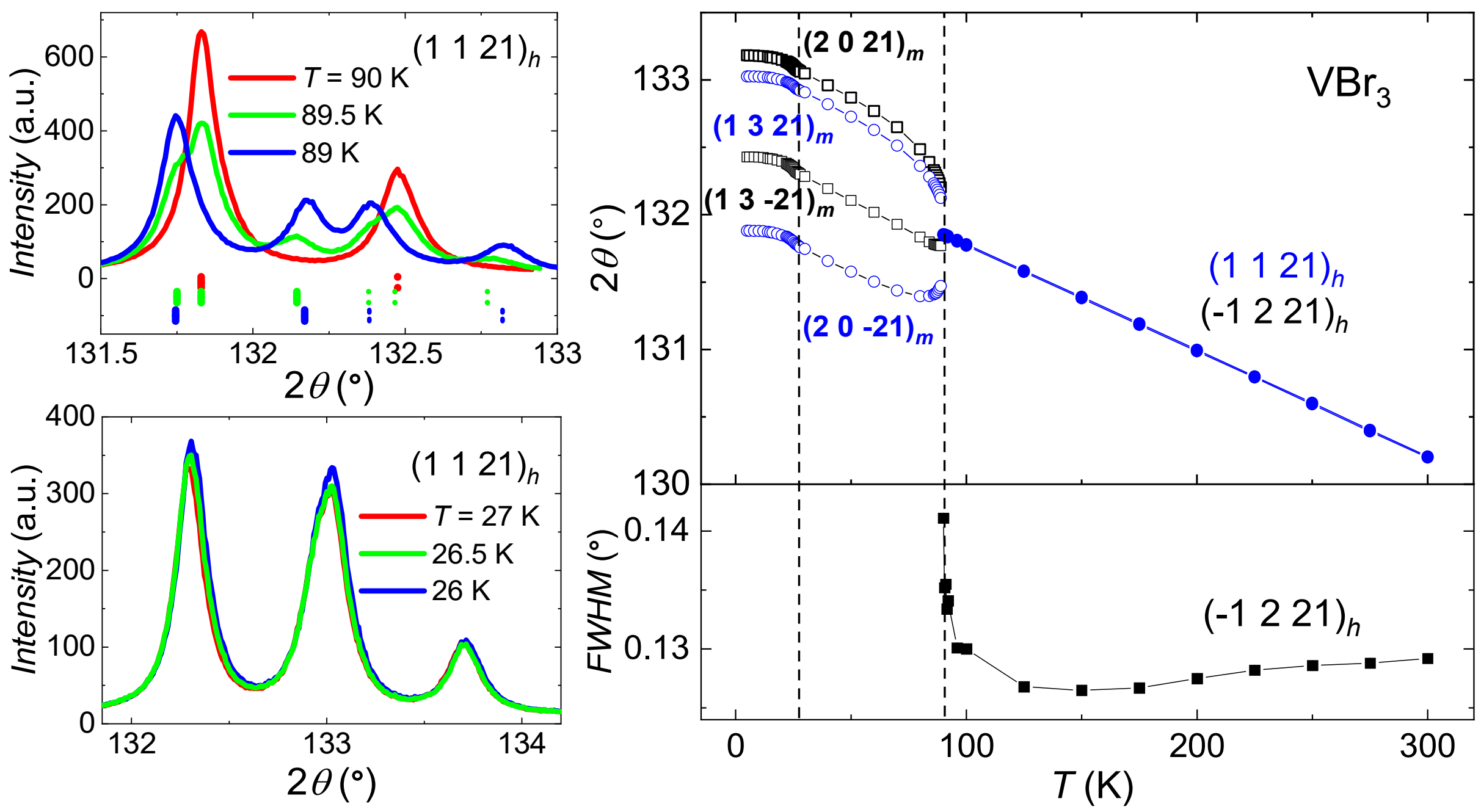

Figure 2. Left panel: Evolution of the (1 1 21) diffraction peak of a $VBr_3$ single crystal in the temperature interval from 89 to 90 K (top) and from 26 to 27 K (bottom). The vertical solid and dotted markers on the upper figure indicate the $Cu_{K\alpha1}$ and $Cu_{K\alpha2}$ positions of the $Cu_{K\alpha1,2}$ doublet, respectively; Right panel: Temperature dependence of $2\theta$ of the (1 1 21) and (-1 2 21) diffraction peaks (upper panel) and the diffraction-peak half-width *FWHM* of (-1 2 21) (lower panel). The vertical dashed lines mark the positions of the structural and AFM phase transitions. The error bars are smaller than the markers.

On the contrary, we do not observe any (H K 21) diffraction-peak splitting at temperatures below 27 K (see lower panel of Figure 2), where the specific heat reveals the existence of an AFM transition. In other words, the LT crystal symmetry is not affected by the AFM transition in the same way as it was affected by the ferromagnetic (FM) transition in $VI_3$. The absence of further peak splitting down to the lowest temperatures suggests, that there is no lowering of lattice symmetry. Temperature dependence of lattice parameters and unit cell volume parameters of $VBr_3$ compared to$VI_3$ are summarized in Figure 3. All lattice parameters and the volume $V$ of both compounds are scaled to be able to compare the changes at the transitions directly. A small drop in the temperature dependence of the $c$-lattice parameter and the step-like change in the $a$, $b$, and $\beta$ lattice parameters imply the transition at $T_s = 90$ K. The small step-like change of the unit cell volume in the vicinity of the structural transition is a sign of a typical first-order type phase transition. Although the step is small, the comparison of temperature trends above and below the transition suggests, that it is not an artifact of an experimental error or subsequent data processing. Noticeably, the change of volume is significantly larger compared to the equivalent data in $VI_3$. The tiny changes in lattice parameters visible at temperatures below $T_N$ are most likely due to spontaneous magnetostriction in the AFM state.

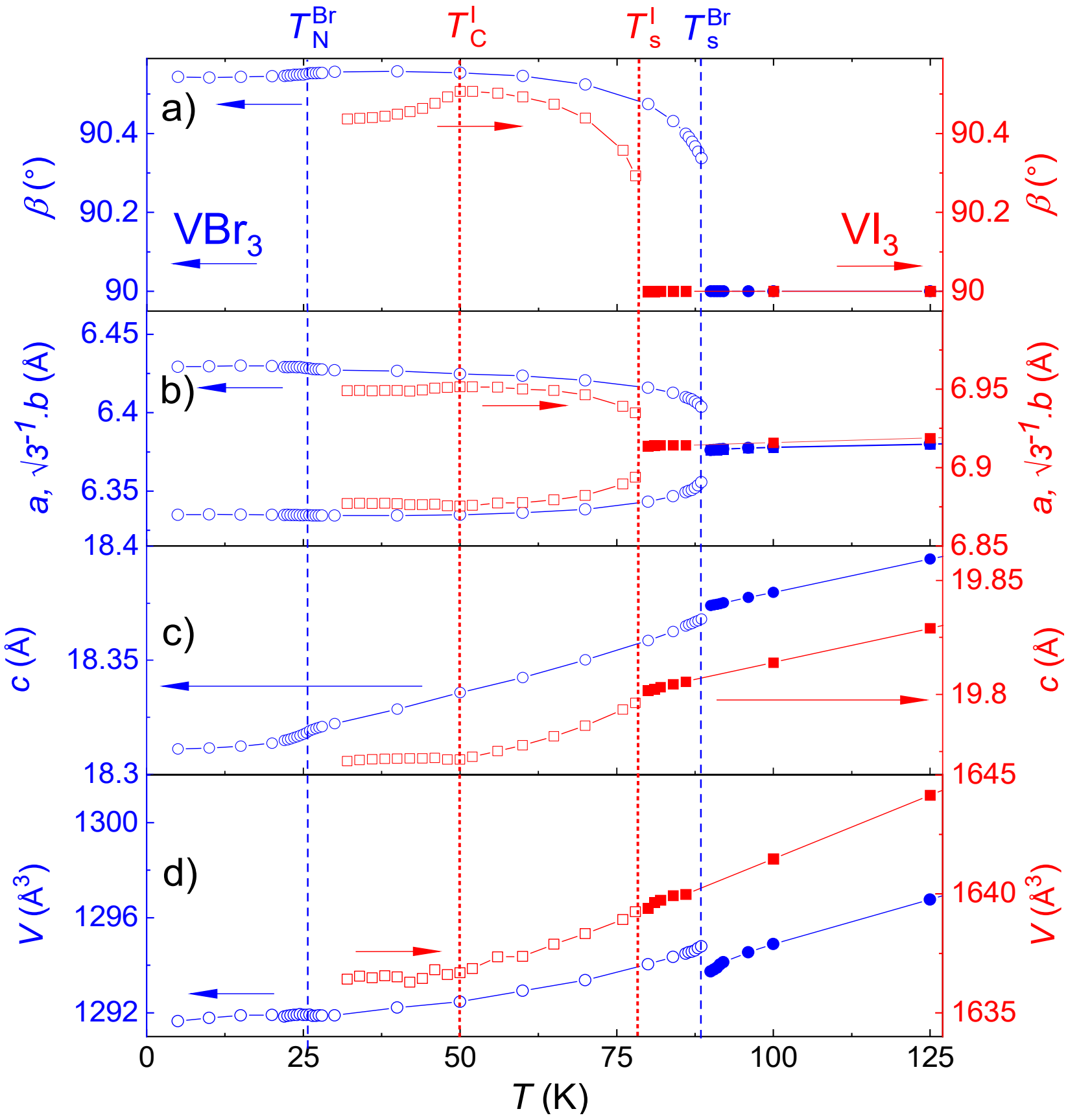

Figure 3. Temperature dependence of lattice parameters and unit cell volume of $VBr_3$ compared to $VI_3$. The left $y$-axis is related to $VBr_3$ while the right $y$-axis is related to $VI_3$ data as depicted by the arrows. The full symbols mark the high-temperature phase while the empty ones mark the low-temperature phase of $VBr_3$ and $VI_3$ data as depicted by the arrow. The vertical dashed and dotted lines mark the magnetic and structural transitions of $VBr_3$ and $VI_3$, respectively.

To track the temperature evolution of diffraction peaks as performed by the single-crystal X-ray diffraction on $VBr_3$ and $VI_3$ [16], we have carried out the experiment and analysis on the $VCl_3$ powder sample. Since the diffraction pattern revealed a strong preferential orientation, i.e. texture, and broadened diffraction peaks, we did not employ the Rietveld analysis as is usual in this type of experiment (for the diffraction pattern, see Figure S1 in Supplementary Mat.). Nevertheless, we managed to index the low-angle part of the diffraction pattern using the structural model used in Ref. [20] and identify the 3 0 0 and 0 0 3 reflections that do not overlap with other peaks. Therefore, they were selected for further study of the diffraction peak splitting as shown in Figures 4 and 5. Upon cooling, the FWHM of the studied diffraction peak increases, and below 100 K, it splits into two peaks; in line with the observation in $VBr_3$ and $VI_3$, both split pairs have different $2\theta$ distances. The reduction of the lattice symmetry is demonstrated by the diffraction-peak splitting with two diffraction peaks moving to larger and the other to smaller $2\theta$ values, respectively. Figure 4 depicts the coexistence of the diffraction peak belonging to the HT structure phase and of the pairs of diffraction peaks from the LT phase which takes place in the temperature region of ~ 10 K, being a significantly larger interval compared to ~ 1 K in $VBr_3$ and $VI_3$.

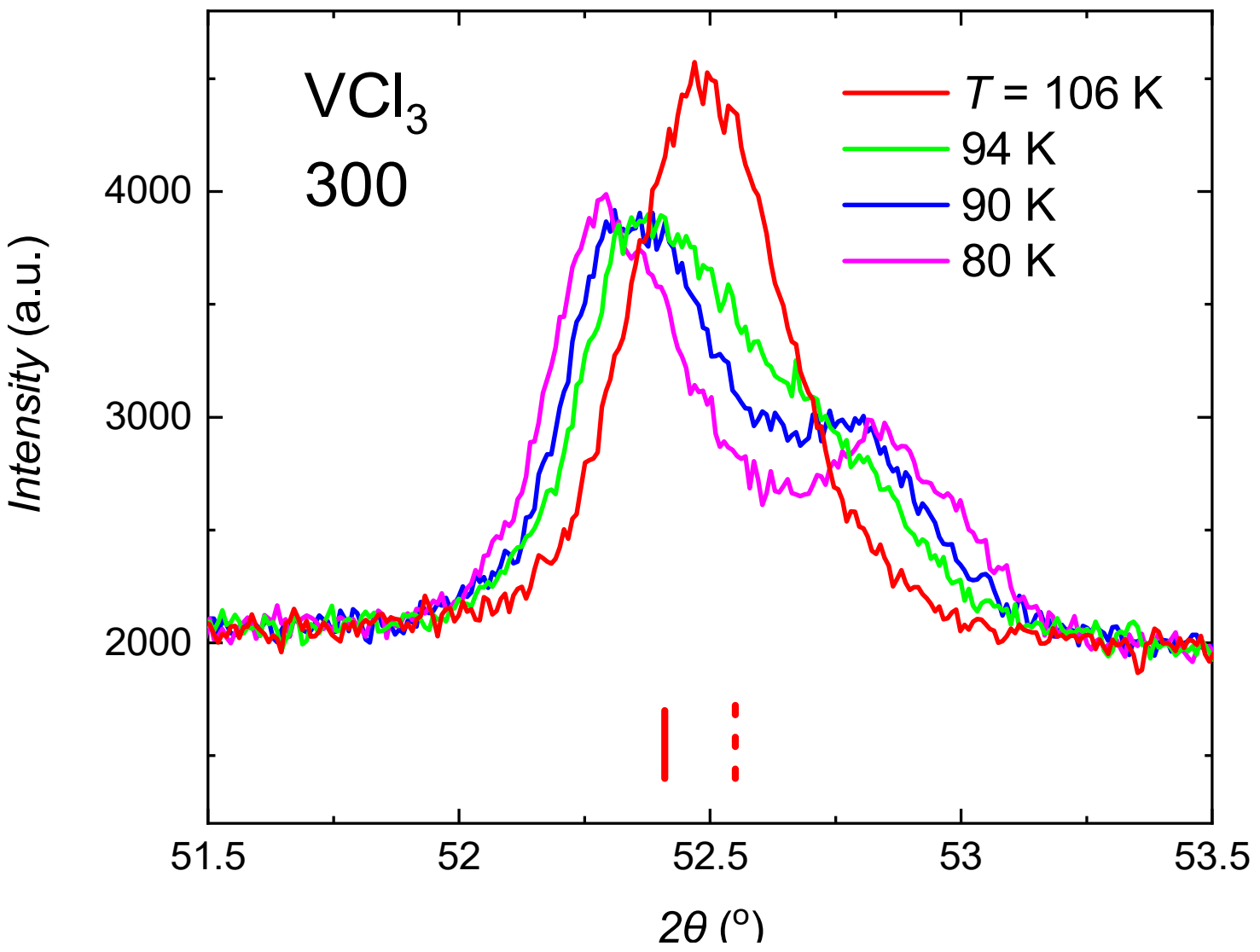

Figure 4. Evolution of the 3 0 0 diffraction peak of a $VCl_3$ powder pattern in the temperature interval from 80 to 106 K. The vertical solid and dotted markers indicate the $Cu_{K\alpha1}$ and $Cu_{K\alpha2}$ positions of the $Cu_{K\alpha1,2}$ doublet.

Comparison of the $2\theta$ temperature dependence of the 0 0 3 peak, the corresponding $d_{001}$ distance, and the specific heat of the polycrystalline sample of $VCl_3$ is shown in Figure 5. The scattered $d_{001}(T)$ evolution is caused by analyzing a diffraction peak with a low $2\theta$-angle.

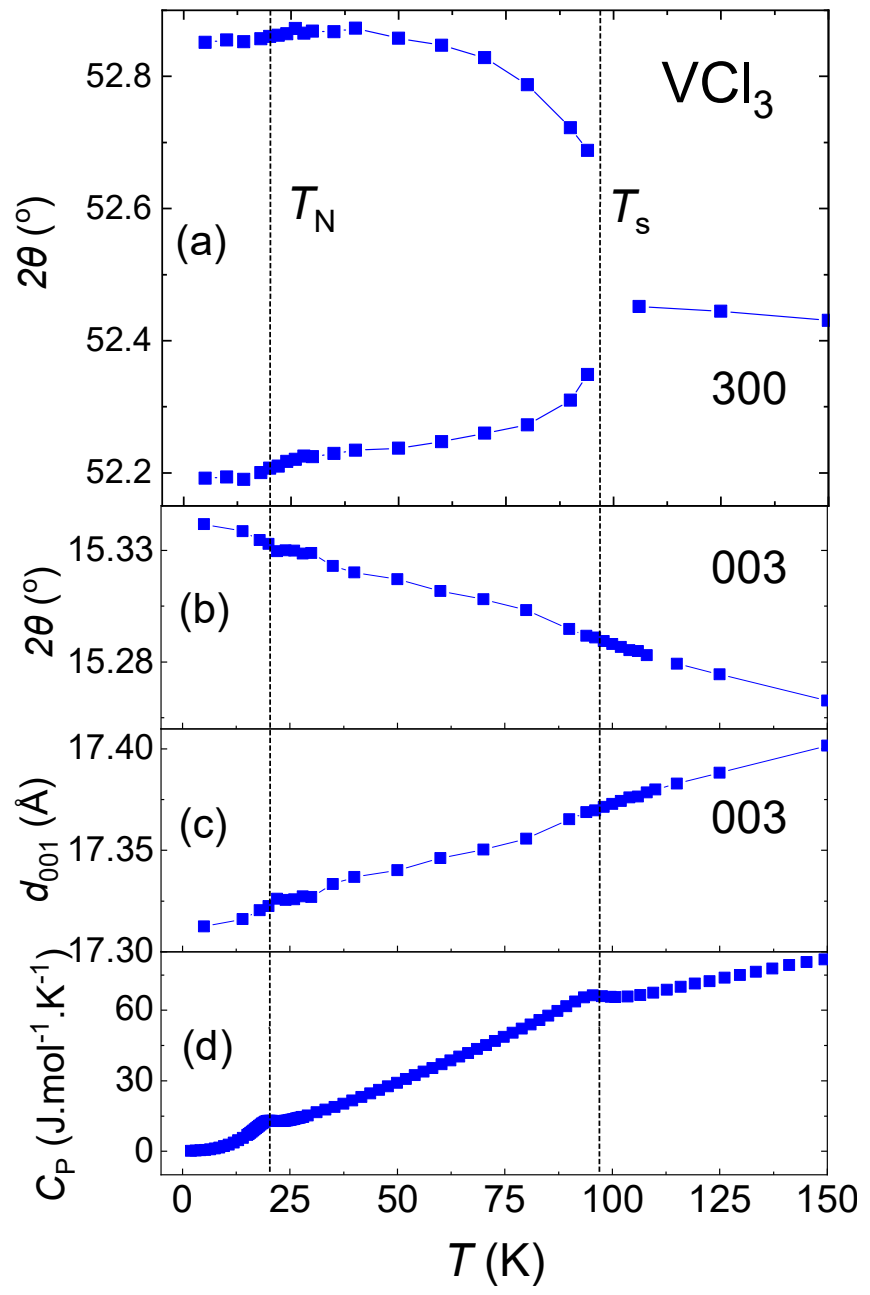

Figure 5. (a) Temperature dependence of $2\theta$ of the 3 0 0 diffraction peak in $VCl_3$, (b) temperature dependence of the $2\theta$ angle of the 0 0 3 diffraction peak, (c) the $d_{001}$ interlayer distance, and (d) specific heat $C_p$ measured at zero magnetic field. The vertical dashed lines mark the positions of the structural and AFM phase transitions at $T_s$ and $T_N$, respectively.

Detailed specific-heat measurements of all studied $VX_3$ ($X$ = Cl, Br, I) compounds are summarized in Figure 6. Despite the shallow peaks of $VCl_3$ caused by its polycrystalline origin, we can see a clear trend in the evolution of the structural and magnetic transitions across the halogen series Cl-Br-I. With increasing atomic number, the structural transition temperature decreases, and the magnetic (AFM in case of Cl and Br, FM in case of I) transition temperature increases. This behavior suggests that the structural and magnetic behavior in these materials is tightly intertwined. While the magnetic field applied perpendicular to the layers tends to suppress the structural transition in $VI_3$ by almost 3 K in the 14 T field [16], it is intact in the Cl and Br compounds (up to 9 T).

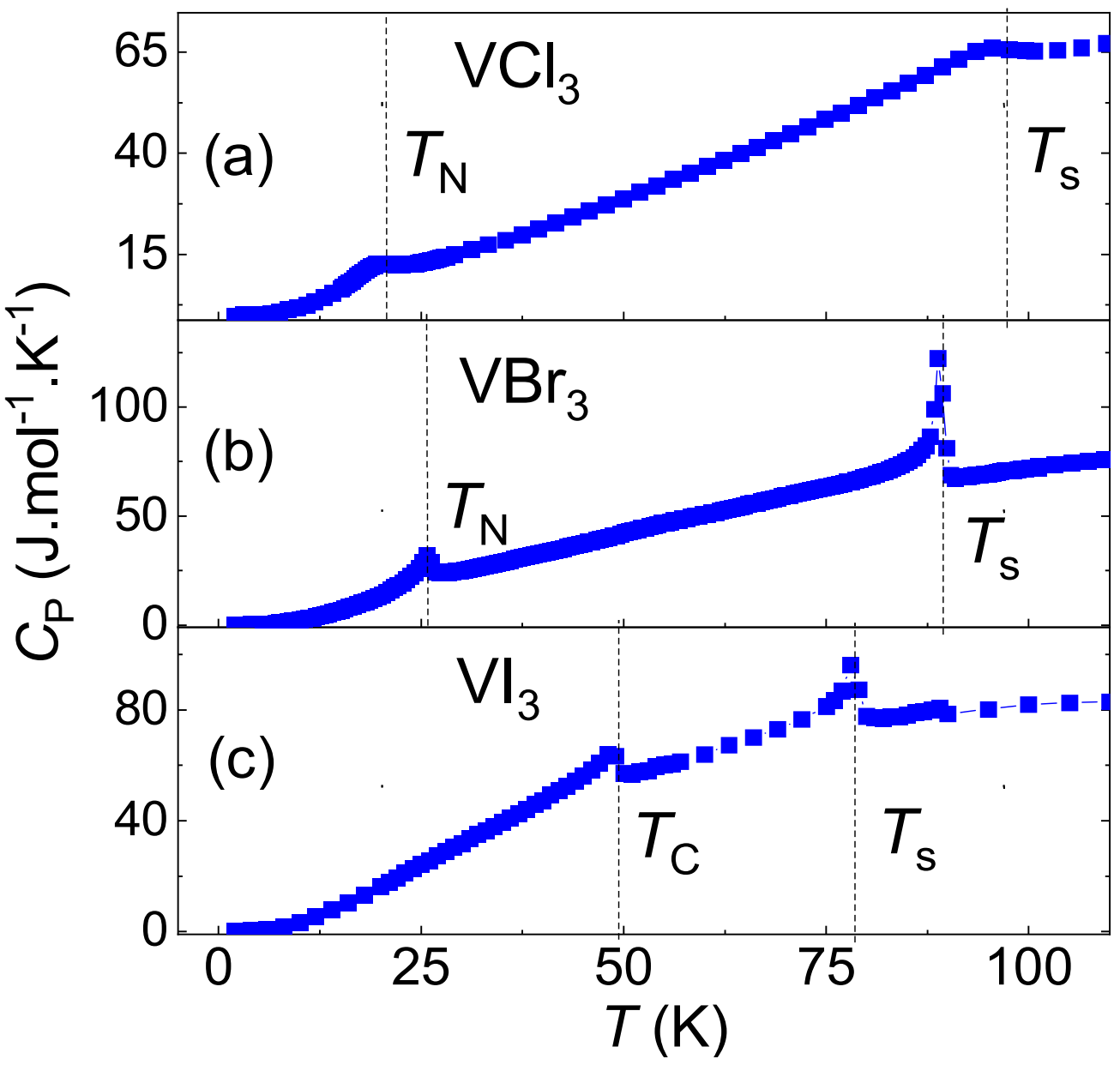

Figure 6. The temperature dependence of the specific heat of the $VX_3$ ($X$ = Cl, Br, I) compounds in zero magnetic field showing the regions with the structural transition at $T_s$ and magnetic transition at $T_m$ (where $T_m$ corresponds to $T_C$ in case of $VI_3$, according to the notation used in Ref. [24], and to $T_N$ in $VBr_3$ and $VCl_3$, respectively).

From the XRD, magnetization, and specific heat data, one can notice four aspects, which differ the behavior of $VBr_3$ and $VCl_3$ from the $VI_3$ system. Primarily, in the first two systems, we observe only one transition related to the peak splitting and thus, symmetry breaking at $T_s$, and second, this structural transition is not magnetic-field dependent contrary to $VI_3$. Next, we observe different magnetic ground states, AFM in $VBr_3$ and $VCl_3$ vs FM in $VI_3$, together with different magnetic anisotropy in the single-crystalline materials. The question of how are these differences related to different inter- and intralayer exchange interactions reflected in stacking faults, dimerization, etc. is a subject of further study.

If we look at Table 2 and related text in McGuire's review [7] and consider the results of Dolezal et al. on $VI_3$ [16], we find that only in the Cr trihalides the HT phase is monoclinic and the LT phase is trigonal whereas in the Ti, V, and Fe analogs possess the trigonal HT phase. Then the $CrX_3$ compounds seem to be exceptional within the family of the 3d-transition metal trihalides concerning the symmetry evolution between these two structural phases. When we inspect corresponding available crystallographic data on them, we find sets for a reasonable analysis only for the Ti, V, Cr, and Fe chlorides and the V, Cr, and Fe bromides. Iodides have been reported only with V and Cr. In Table 1, the values of lattice parameters $a$ and $c$ of trigonal phases of these compounds are listed together with the $c/a$ ratio values. The trigonal phase exists at room temperature except for $CrI_3$ and $CrCl_3$. These two compounds when cooled from high temperatures undergo the monoclinic to trigonal structure transition near 210 and 240 K, respectively. Therefore the parameters of trigonal structure for $CrI_3$ and $CrCl_3$ have been determined at lower temperatures, namely at 225 K for $CrCl_3$ [15] and at 90 K for $CrI_3$ [7] and $VI_3$ [16]. For a better comparison of structure parameters of iodides also data for $VI_3$ determined at the same temperature have been included. When looking at Table 1 shown below we can conclude that the $CrX_3$ compounds exhibit the maximum *c/a* values within chlorides, bromides, and iodides, respectively.

**Table 1.** Lattice parameters *a,c*, and the *a/c* ratio of trigonal phases of several $TMX_3$ compounds at room temperature (225 K for $CrCl_3$; 90 K for $VI_3$, and $CrI_3$).

| Compound | *a* (Å) | *c* (Å) | *c/a* | Reference |
|---|---|---|---|---|
| $TiCl_3$ | 6.12 | 17.50 | 2.86 | [25] |
| $VCl_3$ | 6.01 | 17.34 | 2.89 | [25] |
| $CrCl_3$ | 5.94 | 17.33 | 2.92 | [15] |
| $FeCl_3$ | 6.07 | 17.42 | 2.87 | [26] |
| $VBr_3$ | 6.40 | 18.50 | 2.89 | [20] |
| $CrBr_3$ | 6.26 | 18.20 | 2.91 | [27] |
| $FeBr_3$ | 6.94 | 18.38 | 2.65 | [28] |
| $VI_3$ | 6.92 | 19.81 | 2.86 | [16] |
| $CrI_3$ | 6.87 | 19.81 | 2.88 | [7] |

The unique position of the $CrX_3$ compounds within the $TMX_3$ family based on the reverse structure sequence is further highlighted by the fact that we observe a large thermal hysteresis in $CrX_3$ ($X$ = Cl, I) pointing to the first-order nature of the structural transition. On the other hand, the nature of the structural transition in the $VX_3$ compounds is not manifested so clearly although current data such as the step-like transition in the temperature evolution of $V$ and the possible coexistence of different structure phases bring mild evidence for a first-order transition, as discussed further. No other reliable data regarding the nature of the structural transition in related $TMX_3$ compounds are available so far, but some conclusions can be made already.

Generally, the first-order phase transitions are characterized by the release of the latent heat and a step-like change in the volume. In the case of the second-order phase transition, the volume is a continuous function of temperature and the latent heat is not released. Instead, the discontinuities in specific heat, isobaric thermal expansion, and isothermal compressibility are present. Experimentally, this theoretical classification might not be so obvious, if the volume change and latent heat are very small, which is the case of structural transitions. The $V$(T) of $VI_3$ seems to be an almost continuous function in the vicinity of the transition while the $V$(T) of $VBr_3$ exhibits a very small, but visible step change of the volume. Concerning the presence of latent heat, the situation gets complicated. Its presence leads to the divergence of specific heat while its absence is reflected in a step change and a lambda-type anomaly. However, the measured shape of the anomaly in the specific heat of $VI_3$ and $VBr_3$ does not support the first-order type scenario convincingly. Thus, the step-change in the volume of $VBr_3$ is the only clear indication for the first-order transition. Due to the similarity of transitions in $VI_3$ and $VCl_3$, we can assume that these structural transitions are of the first order, too.

Another character that seems to be typical of these structural transitions is the athermal growth of low- or high-temperature new phases during the transition. The measurement of reciprocal space maps takes around 1.5 hours and during this time the temperature is kept constant. Also, the ratio of both phases seems to be constant and is only changed by the further change of the temperature. This behavior is similar to the one observed in martensitic transformations [29].

## 4. Conclusions

We have demonstrated by the results of measurements of X-ray diffraction, specific heat, and magnetization at various temperatures on $VCl_3$, $VBr_3$, $VI_3$ that these compounds are dimorphic similar to most of the other transition-metal trihalides. The structure of the LT phase is probably monoclinic whereas the structure of the HT phase has a higher symmetry (trigonal $BiI_3$-type in case of $VI_3$). The behavior of $VBr_3$ in the vicinity of the structural phase transition gives several hints that suggest the first-order nature of the transition; however, this scenario has to be confirmed by further experiments. These findings are in striking contrast with the structural behavior of compounds in the corresponding chromium-based triad, $CrCl_3$, $CrBr_3$, $CrI_3$. In these materials, the LT phase has a higher-symmetry trigonal ($BiI_3$-type) whereas the HT phase is monoclinic. The structural transition in $CrCl_3$ and $CrI_3$ is of the first-order type associated with a large thermal hysteresis with the coexistence of both phases. The $CrX_3$ compounds seem to be exceptional within the family of 3d-transition metal trihalides concerning the symmetry evolution between the high- and low-temperature phase. They also exhibit the largest $c/a$ values within the series of chlorides, bromides, and iodides, respectively.

## Acknowledgments

This work is a part of the research project GAČR 21-06083S which is financed by the Czech Science Foundation. The project was partially supported by the OP VVV project MATFUN under Grant No. CZ.02.1.01/0.0/0.0/15_003/0000487. The experiments were carried out in the Materials Growth and Measurement Laboratory MGML (see: http://mgml.eu) which is supported within the program of Czech Research Infrastructures (project no. LM2018096).